\documentclass[conference]{IEEEtran}
\IEEEoverridecommandlockouts
\usepackage{cite}
\usepackage{graphicx}
\usepackage{soul}
\usepackage{color}
\usepackage{xcolor}
\graphicspath{ {./images/} }
\usepackage{amsmath,amssymb,amsfonts}
\usepackage{algorithmic}
\usepackage{graphicx}
\usepackage{textcomp}
\usepackage{xcolor}

\begin{document}

\title{Wireless IoT Energy Sharing Platform}

\author{\IEEEauthorblockN{Jessica Yao, Amani Abusafia, Abdallah Lakhdari, and Athman Bouguettaya}
\IEEEauthorblockA{School of Computer Science\\
 The University of Sydney, Australia\\	
\{jyao7254, amani.abusafia, abdallah.lakhdari, athman.bouguettaya\}@sydney.edu.au}
}

\maketitle

\begin{abstract}
Wireless energy sharing is a novel convenient alternative to charge IoT devices. In this demo paper, we present a peer-to-peer wireless energy sharing platform. The platform enables users to exchange energy wirelessly with nearby IoT devices. The energy sharing platform allows IoT users to send and receive energy wirelessly. The platform consists of  (i) a \textit{mobile application} that monitors and synchronizes the energy transfer among two IoT devices and (ii) and a \textit{backend} to register energy providers and consumers and store their energy transfer transactions. The developed framework allows the collection of a real  wireless energy sharing dataset. A set of preliminary experiments has been conducted on the collected dataset to analyze and demonstrate the behavior of the current wireless energy sharing technology.  

\end{abstract}


\section{Introduction}

The proliferation of the Internet of things (IoT), particularly wearables, may give rise to a self-sustained crowdsourced IoT ecosystem \cite{atzori2010internet}. The augmented capabilities of IoT devices such as sensing and computing resources may be leveraged for peer-to-peer sharing. People can exchange a wide range of IoT services such as computing offloading, hotspot proxies, {\em energy sharing}, etc. These crowdsourced IoT services present a convenient, cost-effective, and sometimes the only possible solution for a resource-constrained device \cite{ahabak2015femto}. For instance, a passenger's smartphone with low battery power may elect to receive energy from nearby wearables {\em using Wifi} \cite{raptis2019online}. The focus of this paper is on crowdsourcing IoT energy services.

\emph{Energy-as-a-Service (EaaS)}, is defined as the wireless delivery of energy from an IoT device (i.e., \textit{provider}) to a nearby IoT device (i.e., \textit{consumer})\cite{lakhdari2020Vision} \cite{lakhdari2020Elastic}. An energy provider refers to an IoT device that can share spare energy. An energy consumer is an IoT device that requires energy. Examples of such IoT devices are smartphones, smartwatches, etc.  Energy providers, such as smart textiles or solar watches, may \textit{harvest} energy from natural resources (e.g. body heat or physical activity) \cite{gorlatova2015movers} \cite{tran2019wiwear} \cite{lu2015wireless}. For instance, the PowerWalk kinetic energy harvester produces 10-12 watts of on-the-move power \footnote{www.bionic-power.com}. The harvested spare energy can be shared with nearby IoT devices as services (hence the name energy-as-a-service). Energy services may be achieved with the emergence of new technologies known as \textit{``Over-the-Air wireless charging''} \cite{abusafia2020incentive}. For example, Energous developed a technology to enable wireless charging up to a distance of 4.5 meters\footnote{www.energous.com}. Another example, Xiaomi’s Mi Air charger sends energy wirelessly to multiple nearby IoT devices\footnote{www.mi.com}.
 
Providing energy services (EaaS) has several advantages: EaaS is a crowdsourced \emph{green} solution as it recycles spare energy \cite{lakhdari2020Vision}. In addition, EaaS offers spatial freedom and convenience as an alternative to carrying power banks or plugging into power outlets \cite{dhungana2020peer}\cite{lakhdari2020fluid} . Moreover, EaaS offers ubiquitous power access where users can charge their devices anywhere and at any time \cite{abusafia2020reliability}\cite{lakhdari2021fairness}. These advantages promote the concept of \emph{wireless crowdcharging} \cite{bulut2018crowdcharging}.

This demo focuses on developing a novel platform to \textit{effectively provision wireless IoT energy services} in a confined area. Our prospective EaaS environment consists of confined areas, also referred to as \emph{microcells}. A microcell is defined by an area where people typically gather; e.g., coffee shops,  movie theaters, and restaurants. In an EaaS environment, IoT devices may offer energy services by sharing their excess energy to neighboring IoT devices. 

\begin{figure}[!t]
    \centering
    \includegraphics[width=\linewidth]{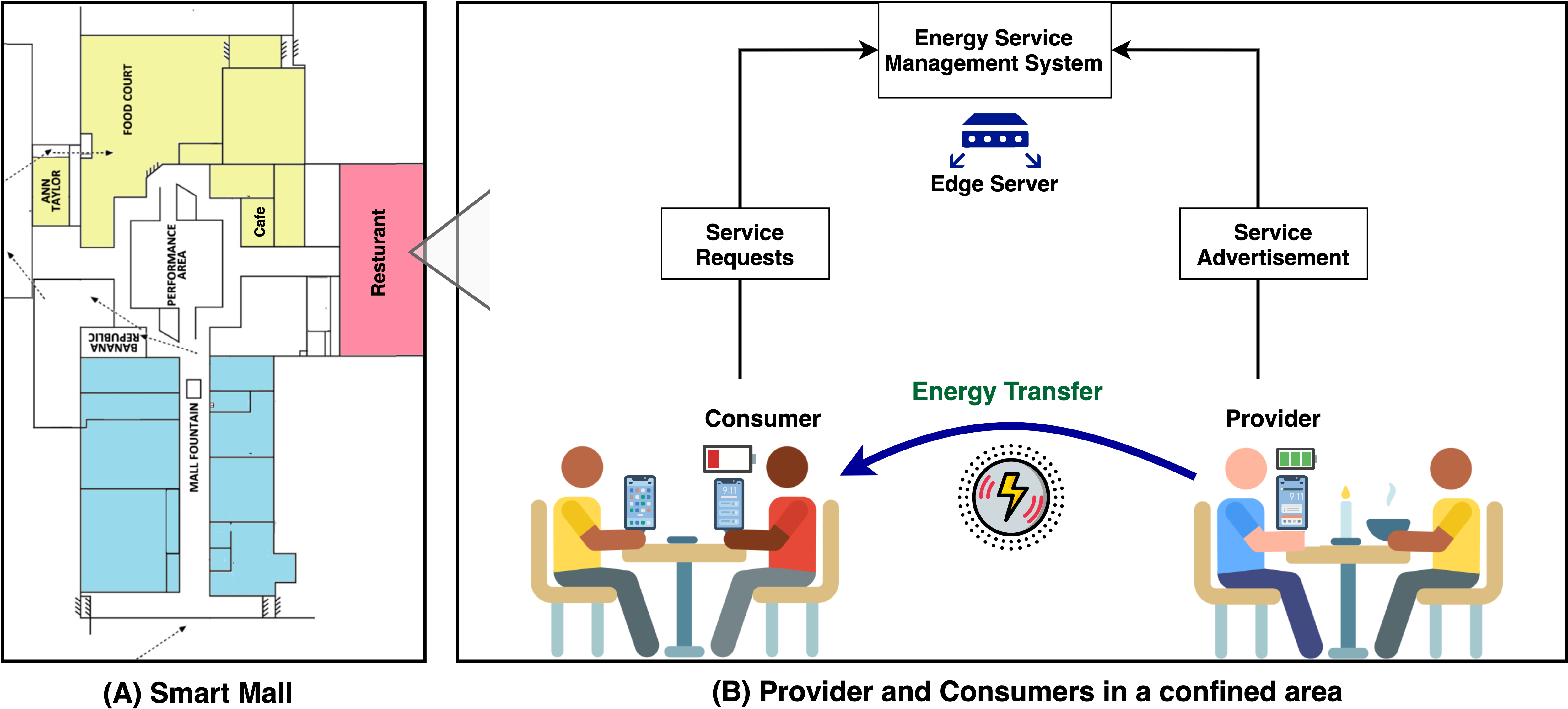}
    \caption{Wireless IoT energy services scenario}
    \label{scenario}

\end{figure}

\section{System Overview}

 We consider the following scenario to illustrate and motivate our demo as shown in Fig.\ref{scenario}. We assume that a mall is split into geographical $microcells$ such as cafes, restaurants, or movie theaters (see Fig.\ref{scenario}A). In a microcell, users own IoT devices such as a smartphone or a smartwatch which act as service providers or consumers (see Fig.\ref{scenario}B). IoT devices may require energy or  have spare energy to share with nearby devices. Users may submit energy services or requests through our developed mobile application. The application will connect users to nearby IoT devices. All local energy services and requests are submitted through the  mobile application to the service management system at the edge e.g., a router associated with the microcell. The mobile application uses Bluetooth to connect energy providers to energy consumers for communication purposes. In addition, the mobile application  synchronizes the monitoring and recording of the energy transfer between the provider and consumer. The current version of the energy sharing application supports one-to-one energy transfer mode, i.e., a \textit{single} energy provider can deliver energy to only a \textit{single} energy consumer.

\begin{figure}[!t]
    \centering
    \includegraphics[width=\linewidth]{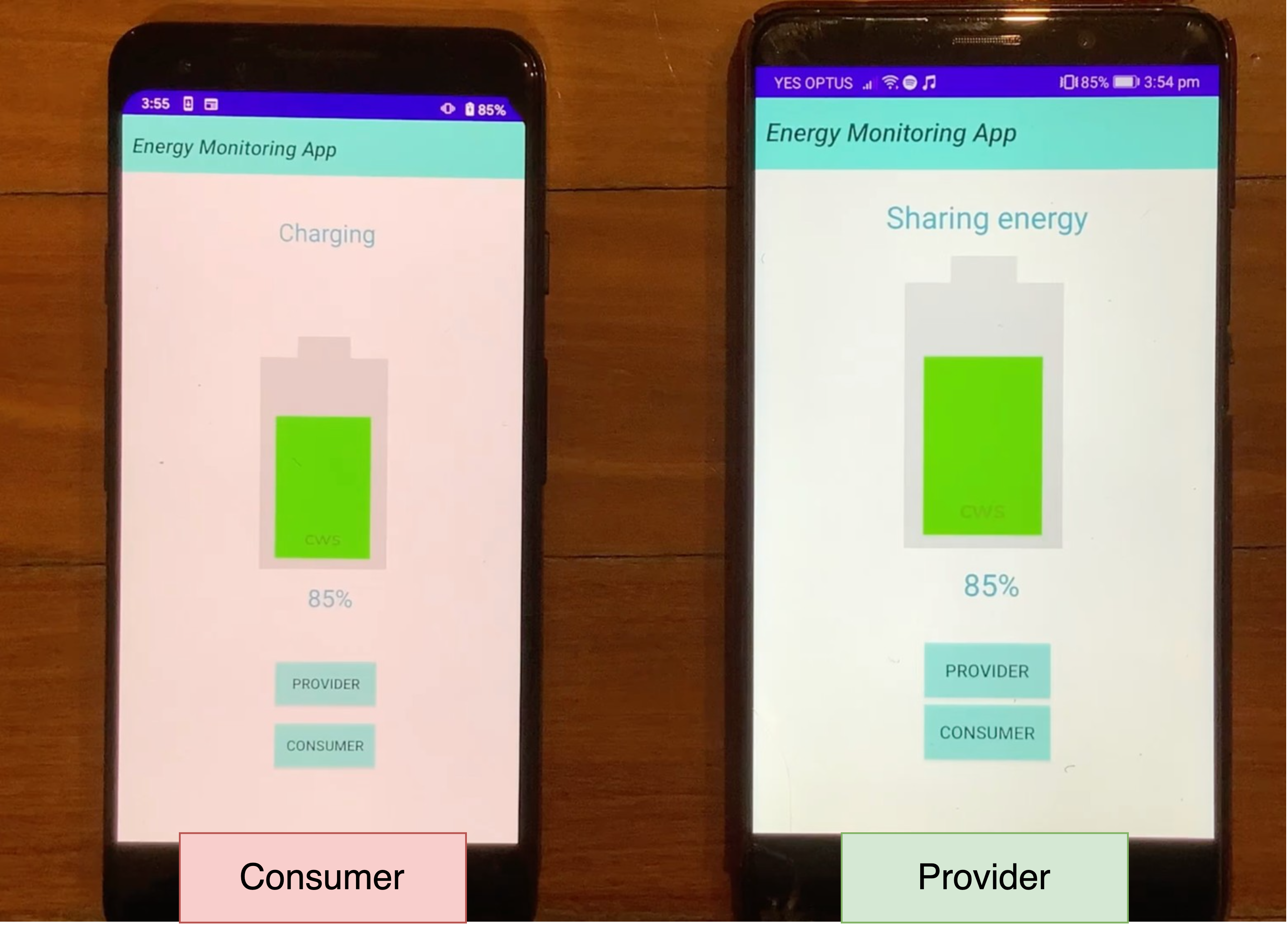}
    \caption{Wireless energy services mobile application}
    \label{SS}
\end{figure}

\section{Demonstration Scenario}
We demonstrate our platform by a real use case of charging a smartphone wirelessly by  another device (See Fig.\ref{SS}). Several companies focus on developing the wireless charging technology of  IoT devices over a distance. Examples of these companies are: Cota\footnote{www.ossia.com}, Powercastco\footnote{www.powercastco.com}, TechNovator\footnote{technovator.co}, and Lumen Freedom\footnote{lumenfreedom.com}.  It is challenging to obtain a prototype from these companies due to the high demand \cite{lakhdari2021fairness}\cite{lakhdari2021proactive}. Therefore, we used the existing technology of wireless charging/recharging in smartphones to demonstrate our platform.  

In this demo, we used the built-in \textit{reverse} wireless charging \textit{transmitter} technology in recent smartphones such as Pixel 5 to act as a provider. We also used the built-in wireless charging \textit{receiver} technology in recent smartphones to act as a consumer.  First, the consumer clicks on the \textit{consumer} button to request energy. On the other hand, the provider clicks on the \textit{provider} button to share energy. Then, both the provider and consumer will specify the amount of offered/required energy as a percentage of their battery. We assume both provider and consumer will offer/request the same amount of energy. The mobile application connects the consumer to the provider using Bluetooth. Note that both devices need to be at a close distance to connect. Once the devices are connected, the consumer sends an energy request to the provider. The provider will be asked to confirm their acceptance to fulfill the consumer's request. If the provider accepts, the consumer will be notified and then will inform the system about the transaction. The consumer then will notify the provider at the start of the wireless energy transfer and send the transaction ID. While charging, the app records the battery status of each device every 5 seconds.  Once the energy request is completed, both provider and consumer will update the transaction record on the backend by sending the battery status log during the energy transfer and their final battery status.

The aforementioned scenario is demonstrated in a demo video. The demo file is titled “Wireless Energy Services Platform". It is an MP4 video format. It can be found in the following link: https://youtu.be/cIvrhJUCq2g

\begin{figure}[!t]
    \centering
    \includegraphics[width=\linewidth]{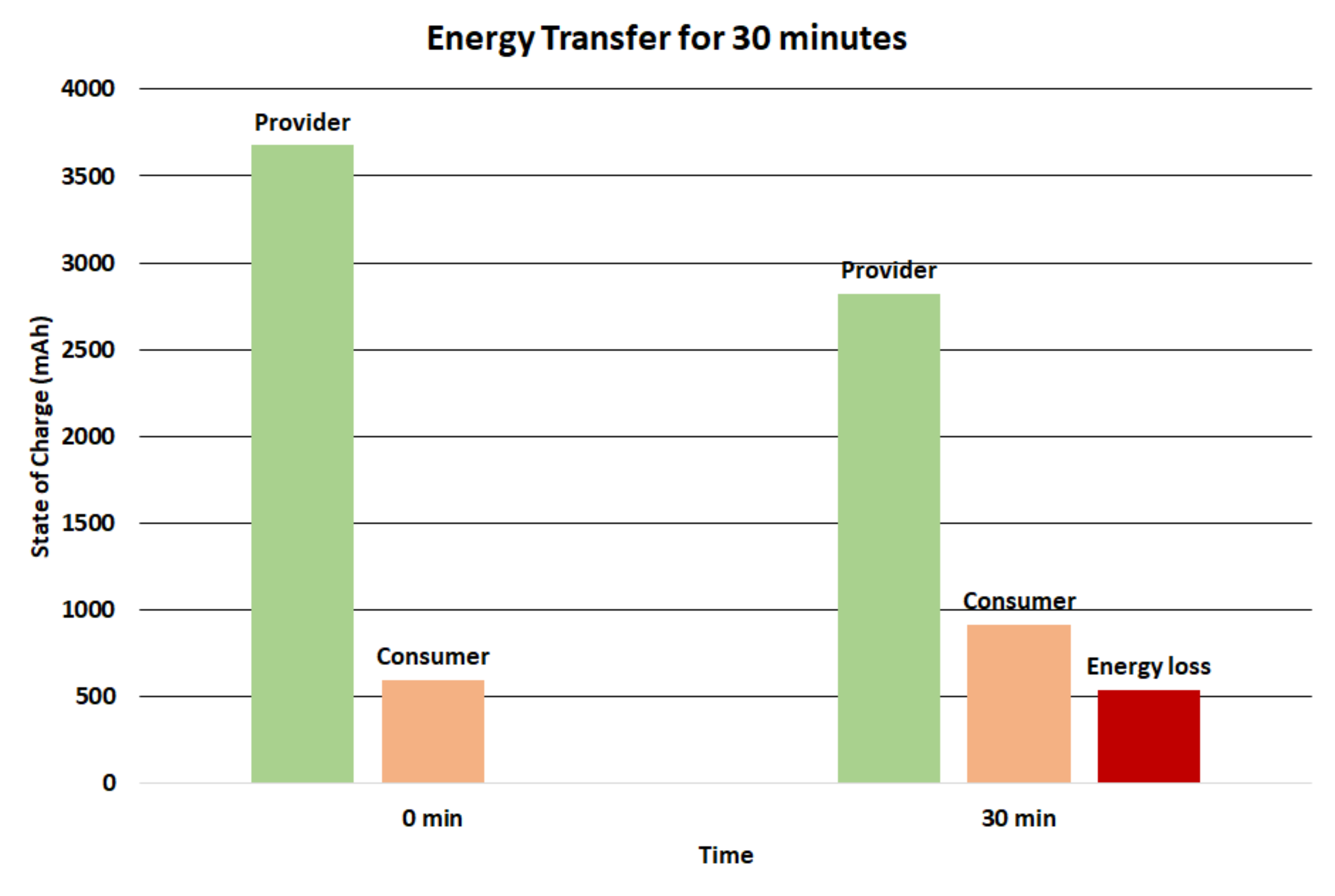}
    \caption{Example of an Energy Transfer for 30 minutes
}
    \label{exp1}
\end{figure}

\section{preliminary experiments}
 We ran a set of preliminary experiments on the collected dataset to analyze and demonstrate the behavior of the current wireless energy sharing technology. A consumer will request energy for a certain time period, i. e. 30 minutes (See Fig.\ref{exp1}). The red column represents the energy loss caused by the wireless energy  transfer.   

\section{PRACTICAL SETUP}
At PERCOM 2022, we will demonstrate the capabilities of our framework in terms of monitoring the wireless energy transfer and pushing the data to the cloud.  At the booth, we will demonstrate the wireless energy sharing process. Interested visitors will also be able to create their energy requests using our mobile application.

\section{Conclusion and Future Work}
In this demo paper, we presented our preliminary implementation of the wireless IoT energy sharing platform. The platform enables energy consumers and providers to connect and establish a wireless energy transfer. The platform uses the current wireless energy charging technology to enable the energy transfer. In the future, we will enable multiple energy services to fulfill an energy request. This may be achieved by composing multiple energy services from multiple providers.

\bibliographystyle{IEEEtran}
\bibliography{main}
\end{document}